\documentclass[sigplan,screen]{acmart}

\renewcommand\footnotetextcopyrightpermission[1]{}

\usepackage[utf8]{inputenc} 
\usepackage[T1]{fontenc} 
\usepackage{hyperref} 
\usepackage{url} 
\usepackage{booktabs} 
\usepackage{amsfonts} 
\usepackage{nicefrac} 
\usepackage{microtype} 
\usepackage{xcolor} 
\usepackage{cite}
\usepackage{algorithm}
\usepackage{amsmath}

\usepackage{amssymb}
\usepackage{mathtools}
\usepackage{amsthm}

\usepackage{multirow}
\usepackage{import}
\usepackage{pifont}
\usepackage{textcomp}
\usepackage{color, colortbl}
\usepackage{graphicx}
\usepackage{subfigure}
\usepackage{makecell}
\usepackage{footmisc}
\usepackage{adjustbox}
\usepackage{tcolorbox}
\usepackage{caption}
\usepackage[noend]{algpseudocode}
\usepackage{bm}
\usepackage[flushleft]{threeparttable}

\usepackage[frozencache,cachedir=.]{minted}

\theoremstyle{definition}

\newcommand{\unishrink}{}
\newcommand{\oursys}{\texttt{FusionLLM}}
\newcommand{\opdata}{\texttt{OP-Data}}
\newcommand{\opnode}{\texttt{OP-Node}}
\newcommand{\ourDAG}{\texttt{OP-DAG}}
\newcommand{\ourscheduler}{\texttt{OP-Fence}}
\newcommand{\ourcompressor}{\texttt{AdaTopK}}

\def\BibTeX{{\rm B\kern-.05em{\sc i\kern-.025em b}\kern-.08em
    T\kern-.1667em\lower.7ex\hbox{E}\kern-.125emX}}
\AtBeginDocument{%
  \providecommand\BibTeX{{%
    Bib\TeX}}}

\copyrightyear{2024}
\acmYear{2024}
\setcopyright{acmcopyright}
\acmConference[Arxiv]{Ungoing}{2024}
\acmPrice{}

\acmPrice{15.00}
\acmISBN{978-1-4503-XXXX-X/18/06}

\begin{document}

\title{\oursys{}: A Decentralized LLM Training System on Geo-distributed GPUs with Adaptive Compression}





\author{
    Zhenheng Tang\textsuperscript{$\dag$}, 
    Xueze Kang\textsuperscript{$\ddag$},
    Yiming Yin\textsuperscript{$\ddag$},
    Xinglin Pan\textsuperscript{$\ddag$},
    Yuxin Wang\textsuperscript{$\sharp$}, 
    Xin He\textsuperscript{$\spadesuit$}, \\
    Qiang Wang\textsuperscript{$\S$},
    Rongfei Zeng\textsuperscript{$\triangle$},
    Kaiyong Zhao\textsuperscript{$\diamondsuit$},
    Shaohuai Shi\textsuperscript{$\S$}, \\
    Amelie Chi Zhou\textsuperscript{$\sharp$}, 
    Bo Li\textsuperscript{$\dag$},
    Bingsheng He\textsuperscript{$\natural$},
    Xiaowen Chu\textsuperscript{$\ddag$}
    }
\affiliation{\institution{
\textsuperscript{$\dag$} The Hong Kong University of Science and Technology  \\
\textsuperscript{$\ddag$} The Hong Kong University of Science and Technology (Guangzhou) \\
\textsuperscript{$\sharp$}Hong Kong Baptist University
\textsuperscript{$\spadesuit$}A*STAR
\textsuperscript{$\triangle$}Northeastern University
\textsuperscript{$\S$}Harbin Institute of Technology, Shenzhen \\
\textsuperscript{$\diamondsuit$}XGRIDS
\textsuperscript{$\natural$}National University of Singapore} \country{\ } \\
\{zhtang.ml, bli, xwchu\}@ust.hk, 
\{xkang507,yyin464,xpan413\}@connect.hkust-gz.edu.cn, \\
\{yxwang,amelieczhou\}@comp.hkbu.edu.hk, hex@cfar.a-star.edu.sg, \\
\{qiang.wang,shaohuais\}@hit.edu.cn, zengrf@swc.neu.edu.cn, \\
hebs@comp.nus.edu.sg, kyzhao@xgrids.com 
}

\renewcommand{\shortauthors}{Z. Tang, X. Kang, Y. Yin, X. Pan, Y. Wang, X. He, Q. Wang, R. Zeng, K. Zhao, S. Shi, A. C. Zhou, B. Li, B. He, X. Chu}

\begin{abstract}

To alleviate hardware scarcity in training large deep neural networks (DNNs), particularly large language models (LLMs), we present \oursys{}, a decentralized training system designed and implemented for training DNNs using geo-distributed GPUs across different computing clusters or individual devices. Decentralized training faces significant challenges regarding system design and efficiency, including: 1) the need for remote automatic differentiation (RAD), 2) support for flexible model definitions and heterogeneous software, 3) heterogeneous hardware leading to low resource utilization or the straggler problem, and 4) slow network communication. To address these challenges, in the system design, we represent the model as a directed acyclic graph of operators (OP-DAG). Each node in the DAG represents the operator (or say layers) in the DNNs, while the edge represents the data dependency between operators. Based on this design, 1) users are allowed to customize any DNN without caring low-level operator implementation; 2) we enable the task scheduling with the more fine-grained sub-tasks, offering more optimization space; 3) a DAG runtime executor can implement RAD withour requiring the consistent low-level ML framework versions.

To enhance system efficiency, we implement a workload estimator for each operator and design an OP-Fence scheduler to cluster devices with similar bandwidths together and partition the DAG to increase throughput. Additionally, we propose an AdaTopK compressor to adaptively compress intermediate activations and gradients at the slowest communication links. To evaluate the convergence and efficiency of our system and algorithms, we train ResNet-101 and GPT-2 on three real-world testbeds using 48 GPUs connected with 8 Mbps $\sim$ 10 Gbps networks. Experimental results demonstrate that our system and method can achieve 1.45 - 9.39 $\times$ speedup compared to baseline methods while ensuring convergence.

\end{abstract}


\maketitle

\section{Introduction}\label{sec:intro}
The evolution of machine learning (ML) has given rise to large language models (LLMs) with vast parameters, such as GPT-3~\citep{gpt,chatgpt,scienceChatGPT}, PaLM~\citep{chung2022scaling}, and DALL-E2~\citep{DALL-E2}, which can mimic diverse linguistic styles and facilitate advanced human-like interactions, symbolizing a significant breakthrough in artificial intelligence~\citep{zhou2023comprehensive,liu2023summary}. The burgeoning parameters and datasets in LLMs necessitate substantial GPU memory and computational power, exceeding the advancements in hardware development. Pre-training a GPT-3 of 175B parameters requires H100 to run at least 13.17 years or RTX 4090 with 60.28 years~\citep{gpt3,EmpComputOptLLM}.

The widening gap between hardware development and the rapid evolution of LLMs presents significant challenges to researchers and engineers without substantial hardware resources. Concurrently, data privacy has emerged as an increasingly prominent concern in recent years~\citep{mcmahan2017communication,european_commission_regulation_2016,tang2022gossipfl,tang2024fusefl,EmpComputOptLLM}. Collecting user data with large companies to train or acquire LLM services can expose user privacy. Such hardware scarcity and privacy concerns motivate us to think about a question: \textit{Is it possible to collaborate with GPUs from multiple geo-distributed persons or parties to train an LLM?}

Some works including SETI@home~\citep{895191} and CoolStreaming~\citep{DONet} aggregate geo-distributed computing and bandwidth resources to complete tasks. These pioneering works bear a striking resemblance to our decentralized training concept for LLMs. In this paper, we propose a decentralized training system, \oursys{}, to train or finetune LLMs through geo-distributed GPUs. \oursys{} splits LLMs into multiple sub-models and deploys them on various geo-distributed devices. Thus, different individuals or entities can contribute their GPUs (called CompNodes) to collaboratively train a LLM, and jointly share the trained model as LLM service. During training, only activations and gradients will be communicated between devices. The raw data is never sent out, thus protecting data privacy to some degree.

However, decentralized collaborative training presents four special critical challenges in terms of system design and efficiency. 
First, current ML frameworks do not support \textbf{remote automatic
differentiation} (RAD) over Internet. 
Second, real-world participants usually have \textbf{different software} including Cuda and ML frameworks. It is difficult to make all participants synchronize the software environments, which is overlooked by several decentralized training works~\citep{ryabinin2023swarm,DeDLOC,yuandecentralized,tang2022gossipfl}. 
Third, different CompNodes may provide \textbf{heterogeneous hardware} in terms of various GPU and CPU architectures, memory sizes, network bandwidth and computing ability. Naive partition and scheduling may result in low resource utility and prolonged communication time.
Last, geo-distributed devices usually communicate via Internet, of which the \textbf{low network bandwidth} (10 Mbps$\sim$10 Gbps)~\citep{yuandecentralized,DecentMOE,tang2022gossipfl} may lead to unacceptable communication time, especially with the substantial amount of data exchanged in LLMs. Thus, the bottleneck of the system throughput (defined as processed samples/second) is dominated by the communication time. It is important to compress the communicated data (intermediate features or gradients) between different devices.

To address above challenges, in our system design, we abstract the model as a directed acyclic graph of operators (\ourDAG{}), which is independent of low-level systematic details and specific ML frameworks. Each layer (operator) in the DNN is as abstracted as \opnode{}. Each directed edge is the forward or backward propagation between layers, in which the output activations and gradients are packaged as \opdata{} communicated according to directed edges. The whole \ourDAG{} is partitioned into sub-DAGs for distribution across multiple GPUs. Based on the \ourDAG{}, we design a runtime executor manages these sub-DAGs and executes forward propagation (FP) and backward propagation (BP) to implement RAD. Thanks to the abstracted RAD implementation, our system can use different ML frameworks as ML engines on local devices.

For system efficiency, we estimate the computational workload of each \opnode{} and the communication time of each \opdata{}, defining the partitioning and allocation of the model as an optimization problem to minimize the training iteration time. Our scheduler, \ourscheduler{}, solves it by intelligently clustering nodes with similar bandwidth and partitioning the \ourDAG{}. To address the low bandwidth problem, we introduce \ourcompressor{}, an adaptive mechanism that selectively compresses intermediate data at the slowest communication points, ensuring system performance while preserving training convergence.

\textbf{Contributions.} We make the following contributions.
(1) We motivate, identify key challenges and optimization opportunities of decentralized training with geo-distributed GPUs (\S~\ref{sec:backgroundmotivation}). 
(2) We design a general decentralized trainig system \oursys{} with a general model definition \ourDAG{}, its executor and unified data structure \opdata{}, to support remote automatic differentiation, flexible model definitions and heterogeneous software (\S~\ref{sec:DecentSystem}).
(3) We implement a workload estimator for each layer (\S~\ref{sec:EstCompComm}), analyze the overall throughput (\S~\ref{sec:throughput}), and design a scheduler to reduce the communication overheads thus increasing system throughput (\S~\ref{sec:opfence}).
(4) We propose a Top-K sparsification compressor \ourcompressor{} to adaptively compress intermediate activations and gradients (\S~\ref{sec:adaTopk-AG}).
(5) We implement \oursys{} (\S~\ref{sec:implementation}) and conduct real-world experiments across three clusters with 48 heterogoneous GPUs connected with 8 Mbps $\sim$ 10 Gbps Internet and Ethernet. Experimental results show that our system and method can outperform baseline methods 1.45-9.39 $\times$ speedup (\S~\ref{sec:exp}).

\section{Background and Motivation}\label{sec:backgroundmotivation}


\subsection{LLM Training}\label{sec:LLMtraing}

\begin{table}[t!]
\caption{Comparing different GPUs to pre-train GPT-3 (175B). The FLOPS of pre-training the GPT-3 is 3.14E+23~\citep{gpt3}. The column ``GPU days'' shows how many days the according GPU need to run to train a GPT-3. The last column shows the number of GPUs needed to load a GPT-3 in GPU memory. Considering the memory of input, activations and gradients, the actual necessary number of GPUs are more than shown ones. Prices of GPUs are searched lowest prices on \url{https://www.amazon.com/} by 10th October 2023.}
\begin{center}
\resizebox{\linewidth}{!}{
\begin{tabular}{lccccc}
\toprule
\multirow{2}{*}{GPU} & \multirow{2}{*}{Price} & \multirow{2}{*}{TFLOPS} & GPU  & \multirow{2}{*}{Memory} & \# GPUs to \\
&  &  & days &  & load GPT-3 \\
\midrule
H100 & \$37,799 & 756 & 4807   & 80GB &  9  \\
A100 & \$6,780 & 311.84 &  23308 & 80GB & 9  \\
RTX 4090 & \$1,699  & 165.16 & 22004 & 24GB & 30  \\
RTX 4080 & \$989 & 97.5 & 37274 & 16GB & 44  \\
RTX 3080 & \$679 & 59.5 &  61079 & 10GB & 70 \\
\bottomrule
\end{tabular}
}
\end{center}
\label{tab:GPUPerf}
\end{table}

A typical loop of training Deep neural networks (DNNs) consists of loading and passing inputs layer by layer to generate outputs. A loss function is defined to measure different between outputs and actual data labels. Then, the autograd is conducted to automatically obtain gradients on parameters (weights) of DNNs. At last, parameters are adjusted based on gradients with some optimizers like Stochastic Gradient Descent (SGD). 

The training of contemporary LLMs imposes immense computational and memory requirements. For instance, OpenAI's GPT-3~\citep{gpt,chatgpt,scienceChatGPT}, a model that has set milestones in natural language understanding and generation, consists of a staggering 175 billion parameters. This sheer size necessitates not just vast computational power, but also immense memory capacity to store intermediate activations, weights, and gradients during training. We compare different GPUs in training a GPT-3 with 175B parameters in Table~\ref{tab:GPUPerf}, which shows that pre-training GPT-3 requires H100 to run at least 13.17 years or RTX 4090 with 60.28 years. This trajectory underscores the pressing need to explore innovative methods to efficiently and feasibly train such behemoths.

\subsection{Distributed Training}\label{sec:Dist}
\begin{figure}[t]
\unishrink{}
    \subfigtopskip=2pt
    \setlength{\belowdisplayskip}{2pt}
    \setlength{\abovedisplayskip}{-5pt}
    \subfigbottomskip=2pt
    \subfigcapskip=1pt
   \centering
    {\includegraphics[width=0.99\linewidth]{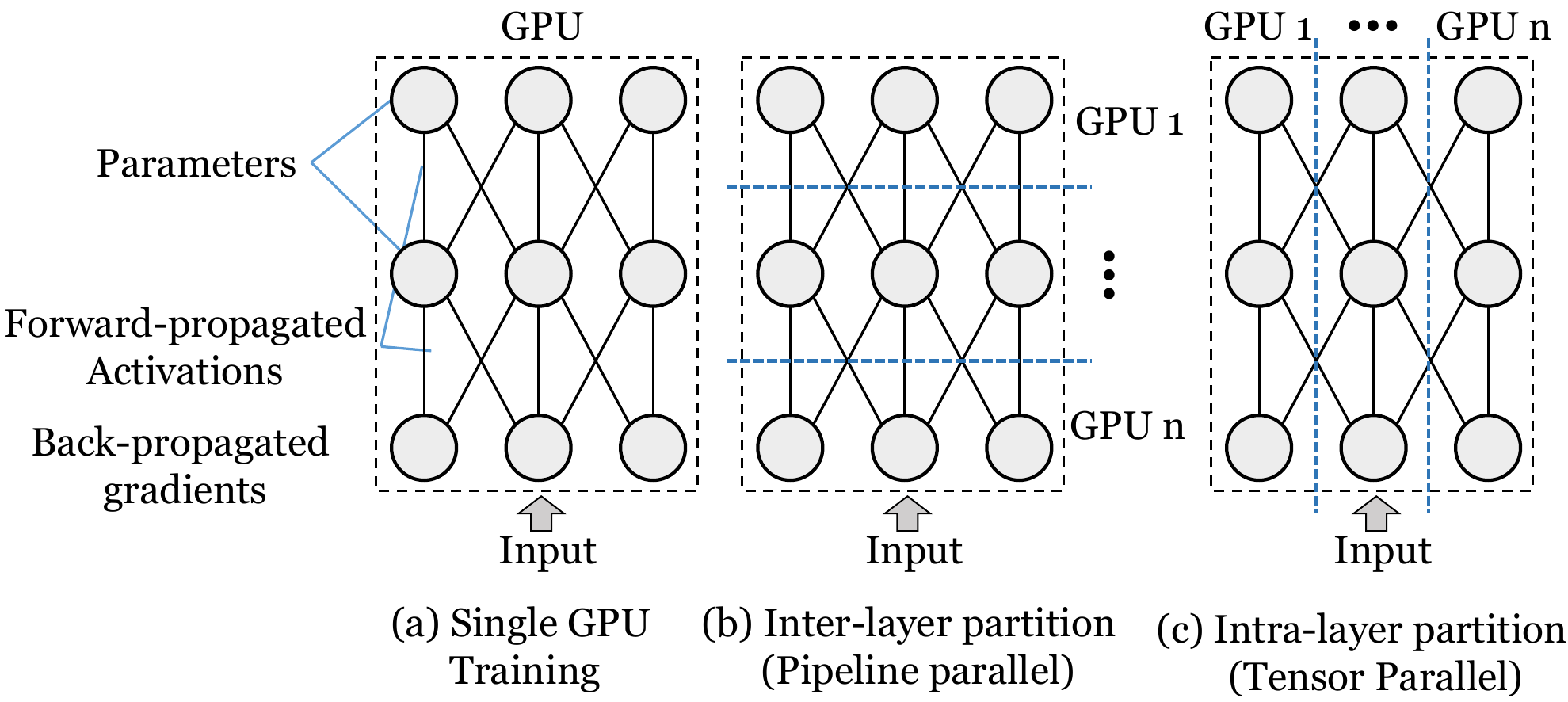}}
    \unishrink{}
    \caption{Training DNNs using single GPUs or distributed GPUs in different modes. Each black line represents activations and gradients when forward propagation and backward propagation respectively.}
    \label{fig:DistributedDNN}
\unishrink{}
\end{figure}

Traditional training and inference of DNNs are distributed within a single organization via high-speed local area network (LAN). In \textbf{Data parallelism (DP)}, input data is divided into subsets and processed on duplicated models across different machines, with the new gradients or parameters from these models being aggregated to update the original model~\citep{tang2020communication,shi2019distributed,tang2024fedimpro,tang2020communication}. However, the scalability of DP is limited by the inefficiency of large-batch SGD~\citep{mccandlish2018empirical}, high communication costs of the whole model size~\citep{tang2020communication,9155269,tang2023fusionai}, and the inability to load an LLM onto a single GPU. Figure~\ref{fig:DistributedDNN} shows two different model parallel schemes, which are widely used when a DNN is too large to be loaded in a single GPU. \textbf{Pipeline parallelism (PP)} dissects model layers into multiple stages and executes them sequentially on different devices, communicating intermediate results between stages~\citep{huang2019gpipe}. Improving the pipelines between execution and communication can reduce bubble time and improve overall efficiency~\citep{narayanan2019pipedream}. However, the layer-wise dependency of forward and backward processes limits the scalability of PP~\citep{park2020hetpipe,tang2023fusionai}.
\textbf{Tensor parallelism (TP)} vertically splits model stages across multiple devices, with each group aggregating computing results and sending them to the other group for the next model stage. TP is suitable for high homogeneous communication-bandwidth environments~\citep{narayanan2021efficient}, because its more fine-grained computation and communication requires perfect overlapping, and introduces higher frequency of communication.

\subsection{Motivation, Challenges and Opportunities}\label{sec:challenge_opportunity}
\textbf{Motivation: aggregating GPUs from multiple individuals or parties.}  Table~\ref{tab:GPUPerf} shows that consumer-level GPUs like RTX 4090 might have a much higher GPU days/price ratio than data-center GPUs like H100. There exists an untapped reservoir of computational power in the form of decentralized GPUs, distributed across various locations and devices. These GPUs, often underutilized, can be harnessed collectively to collaboratively train large models. GPU providers may be motivated to contribute their GPUs for several reasons, ranging from financial incentives to sharing usage rights of LLM and environmental considerations. By decentralizing the training process and capitalizing on these dispersed resources, we not only democratize access to high-end model training but also potentially reduce costs. Instead of investing heavily in dedicated clusters, leveraging the collective power of decentralized GPUs could offer a more economical and scalable solution for training the next generation of AI models.

\textbf{Challenge 1: supporting remote automatic differentiation and the general system design.} Current ML frameworks lack supports for \textbf{automatic differentiation} across Internet networks, hindering seamless remote computational graph processing and gradient calculations. Additionally, the heterogeneity in \textbf{software environments} among real-world participants complicates matters further; variations in Cuda versions and ML frameworks across different nodes make it tedious to synchronize software environments, as a crucial aspect often neglected in existing decentralized training works~\citep{yuandecentralized,ryabinin2023swarm,cocktailSGD,tang2022gossipfl,tang2023fusionai}. 


\textbf{Challenge 2: heterogeneous hardware performance.} 
A pivotal challenge distinct from traditional data center is the inherently heterogeneous hardware landscape of decentralized systems. In such an environment, CompNodes contribute diverse hardware configurations that span a spectrum of GPU and CPU architectures, memory capacities, network bandwidths, and overall computational power. This heterogeneity manifests in variable task completion times across devices. This variability and the strong dependencies during forward and backward propagation of sub-models, result in the "straggler problem"—wherein certain devices lag behind, causing extended waiting periods and potentially throttling the collective training throughput. Addressing this requires meticulous model partitioning strategies, ensuring sub-models are judiciously allocated to devices in a manner that mitigates the straggler effect and optimizes overall system performance.


\textbf{Challenge 3: low network bandwidth.}  
Decentralized training of models across geographically dispersed devices necessitates communication over the Internet~\citep{tang2022gossipfl,tang2020communication}. However, one of the inherent challenges in such a setup is the often low network bandwidth typical of many Internet connections. This limited bandwidth can dramatically escalate the communication time, becoming particularly problematic when considering the voluminous data exchanges inherent to training, such as weight updates or gradient sharing. Consequently, the system's throughput, quantified as processed samples per second, can be critically bottlenecked by this protracted communication duration. The prominence of this challenge underscores the imperative need for innovative strategies to mitigate its impact. One promising avenue is the compression of communicated data—whether it be intermediate features or gradient values—thereby reducing the data payload and aiming to alleviate the prolonged transmission times that can hamstring decentralized training endeavors.

\textbf{Opportunity 1: heterogeneous inter-layer partition.}
Amidst the challenges posed by diverse and heterogeneous hardware configurations in decentralized training, heterogeneous inter-layer DNN partitioning emerges as a promising solution. First and foremost, it enables adaptive load balancing, ensuring that devices, regardless of their varying computational capacities and architecture specifics, receive workload allocations commensurate with their capabilities. This tailored allocation optimizes device utilization, ensuring that no single device becomes a bottleneck due to overburdening or underutilization. Additionally, heterogeneous DNN partitioning offers a strategic advantage over intra-layer partitioning by curtailing frequent communication as shown in Figure~\ref{fig:DistributedDNN}. In the context of an Internet environment, characterized by higher latency per message, intra-layer partitioning can exacerbate communication costs, making the training process inefficient and cost-prohibitive. By adopting inter-layer partitioning, we can judiciously minimize these communication touchpoints, thereby streamlining data exchanges and reducing the latency-induced overheads that can stymie decentralized training performance.


\textbf{Opportunity 2: sparsification property of deep learning.}
Deep learning models exhibit intrinsic properties of sparsification that hold significant implications for both computational efficiency and model robustness. To begin with, these over-parameterized DNNs possess an abundance of parameters, many of which can be pruned or quantized into fewer bits without compromising model accuracy~\citep{frankle2018the,hu2021lora,tang2019doublesqueeze,li2022on,zhang2018three,dong2024pruner}. This property is extensively harnessed for dual purposes: communication compression~\citep{lin2018deep,Hispeed,GCCharge,li2022on,tang2020communication}, where it aids in reducing data transfer volumes, and the acceleration of inference, where it optimizes the real-time performance of deployed models. On another front, the concept of dropping out intermediate activations has been widely adopted as a regularization strategy, staving off the detrimental effects of over-fitting~\citep{Dropout}. Such inherent sparsity in activations and, by extension, gradients, offers an inspiring insight: the potential to further sparsify these intermediate values. This becomes especially crucial in decentralized setups with constrained network bandwidth, where sparsified activations and gradients can substantially curtail communication time, making the training process more agile and resilient to slow network conditions.

\section{\oursys{} Overview}\label{sec:DecentSystem}
\subsection{Design Goals}\label{sec:DesignGoals}

\textbf{General design to support remote automatic differentiation for different ML softwares.} In the fast-evolving landscape of ML, the dynamism and diversity of ML frameworks have become a hallmark of the field~\citep{abadi2016tensorflow,torch}. A system's generality to support the ML framework with different versions to lesser-known but equally significant ones can help CompNode flexibly and swiftly join the computing. By implementing RAD, users do not need to care for the systematic details and focous on the ML aspects. This general design ensures that \oursys{} remains agnostic to the ever-changing preferences in the ML community, offering broader appeal and utility.



\begin{figure}[t!]
    \subfigtopskip=2pt
    \setlength{\belowdisplayskip}{2pt}
    \setlength{\abovedisplayskip}{-5pt}
    \subfigbottomskip=2pt
    \subfigcapskip=1pt
   \centering
    {\includegraphics[width=0.95\linewidth]{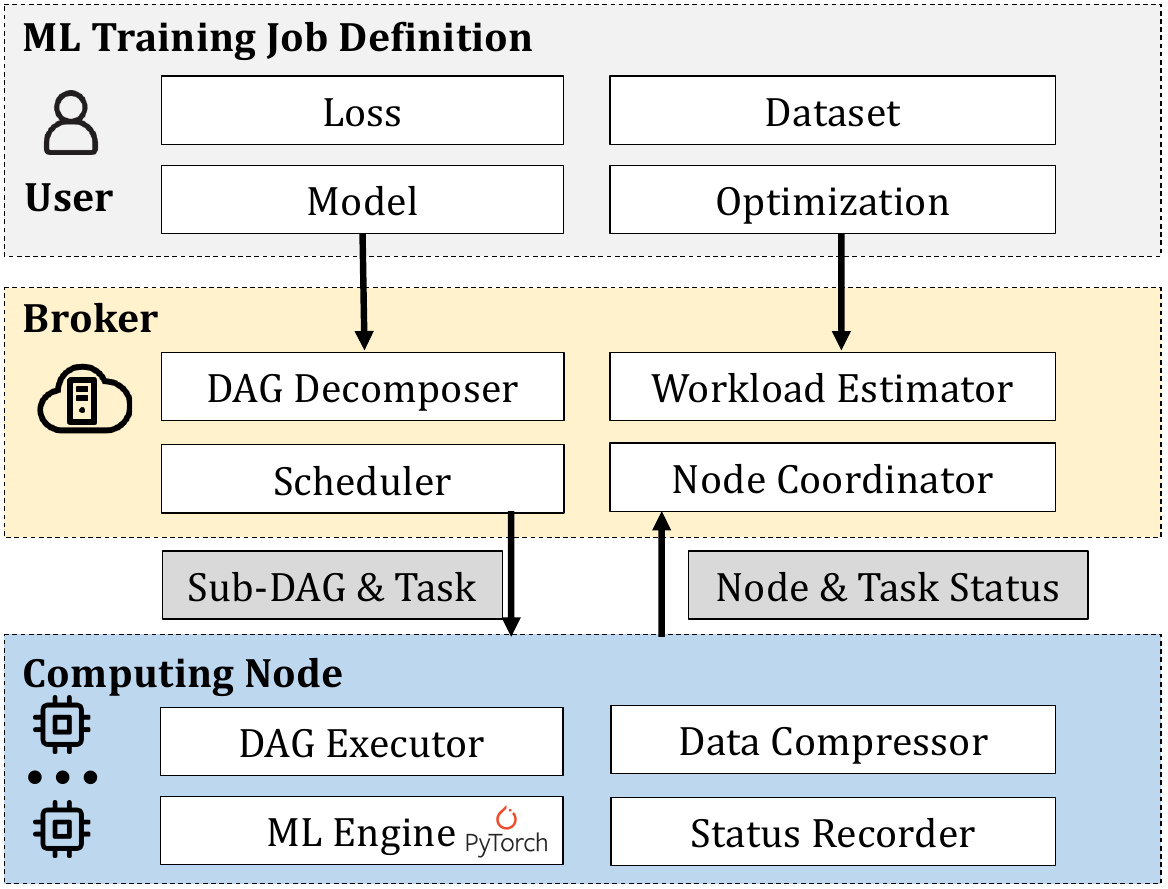}}
    \vspace{-0.3cm}
    \caption{The overview of system design of \oursys{}.}
    \label{fig:systemdesign}
\vspace{-0.3cm}
\end{figure}

\textbf{Compatible design to support convenient customized model architectures.} The world of ML is replete with customized models tailored to address specific tasks or challenges. A system's flexibility in accommodating these customized models, irrespective of their unique architectures or requirements, ensures that it caters to a vast audience ranging from industry professionals to academic researchers. In essence, a general design underscores a commitment to inclusivity and future-readiness, placing the system at the forefront of innovation and ensuring its relevance in the ever-progressing deep learning epoch.

\textbf{Flexible computation scheduling and communication compression.} Many other decentralized frameworks define model architectures coupled with computation and communication codes~\citep{yuandecentralized,ryabinin2023swarm,cocktailSGD}. This rigidity hampers the ability of scheduling algorithms to adaptively allocate workloads, often relegating the task to manual estimation every time a new workload arises, a process that is not just tedious but also prone to inaccuracies. Such a design choice can, inadvertently, magnify the system's operational overhead. Furthermore, as a critical feature for efficient data transfer and storage, communication compression becomes imperative to discern. Without clear demarcation, there is a risk of indiscriminate compression, potentially compromising vital data. These intricacies, when combined, can escalate the code complexity for every model incorporated into the system. Therefore, the emphasis on flexible workload scheduling coupled with judicious communication compression isn't merely about optimizing operations; it is about ensuring a sustainable, efficient, and user-friendly system environment.


\subsection{System Overview}\label{sec:systemoverview}
\begin{figure}[t!]
    \subfigtopskip=2pt
    \setlength{\belowdisplayskip}{2pt}
    \setlength{\abovedisplayskip}{-5pt}
    \subfigbottomskip=2pt
    \subfigcapskip=1pt
   \centering
    {\includegraphics[width=0.8\linewidth]{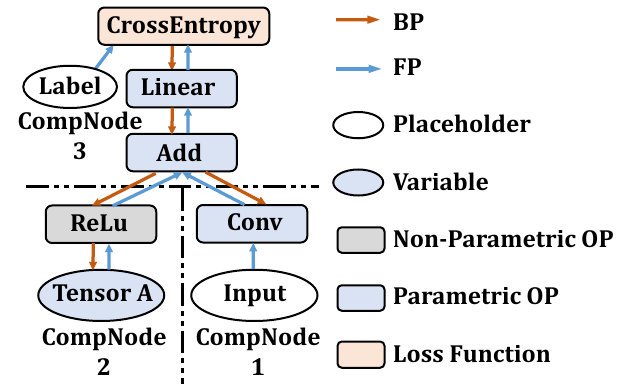}}
    \vspace{-0.1cm}
    \caption{An example DAG of FP and BP processes. CrossEntropy (CE) as the loss function.}
    \label{fig:DAG}
\end{figure}
The system design of \oursys{} is shown in Figure~\ref{fig:systemdesign}. We abstract the definitions of models, datasets, forward functions, loss functions, optimization methods, and training algorithms into the \textit{intermediate representation (IR) plane}. Users only need to provide the definitions to a broker in the IR plane, without need to care about the executing details and scheduling in decentralized training. The broker is designed as a coordinator to receive job definitions, then builds the FP and BP DAG (in \S~\ref{sec:DAG_partition}) of the whole ML models. Each layer (operator) is abstracted as an OP node in DAG. The input, output, and gradients of each layer are saved in a unified data structure (in \S~\ref{sec:datastructure}) to be communicated between OP nodes and devices.

With the DAG, model definitions and parallel training information, we design a scheduling algorithm to automatically partition and allocate different parts of the model as sub-DAGs onto different computing devices. 

Based on the estimation of computing, communication and hardware performance (in \S~\ref{sec:EstCompComm}), the broker estimates system workloads and throughput (in \S~\ref{sec:throughput}) and solves the optimization problem to (in \S~\ref{sec:opfence}). Then, the configurations of partitioned sub-DAGs are sent to allocated computing nodes (CompNode). Each CompNode builds sub-models based on sub-DAGs, and executes FP, BP, and communication operations. For the slowest links indicated by the broker, data communication will be compressed to reduce communication time (in \S~\ref{sec:adaTopk-AG}).

When executing training or inference, we abstract the computing of operators in the DAG, message passing and communication between operators and computing nodes into the \textit{execution plane}. The execution plane is responsible for designing and implementing general interfaces to adapt different ML Engines to execute each operator defined DAGs and communicate data. Thus, the CompNodes can utilize heterogeneous devices and different ML frameworks according to their preference. And there is a DAG execution engine that is responsible for loading and feeding data between different operators.
\begin{table}[t!]
\caption{OP nodes and their attributes in the example DAG (Figure~\ref{fig:DAG}).}
\begin{center}
\begin{small}
\begin{adjustbox}{max width=\linewidth}
\begin{tabular}{lcccc}
\toprule
OP & OP & \multirow{2}{*}{Type} & \multirow{2}{*}{Args}  & CompNode  \\
names & users &  &  & location  \\
\midrule
Input & Conv & Placeholder & -  &  1 \\
Conv & Add & Parametric OP & Input & 1 \\
Tensor A & ReLu & Variable & -  & 2 \\
ReLu & Add & Non-Parametric OP & Tensor A & 2 \\
Add & Linear & Non-Parametric OP & ReLu, Conv & 3 \\
Linear & CE & Parametric OP & Add & 3 \\
Label & CE & Placeholder & -  & 3 \\
CE & - & Loss Function & Label, Linear & 3 \\
\bottomrule
\end{tabular}
\end{adjustbox}
\end{small}
\end{center}
\label{tab:OPNodes}
\end{table}

\begin{table}[t!]
\caption{Sub-graphs and what they require and send in the example DAG (Figure~\ref{fig:DAG}). Acti. means activations during FP, and Grad. means Gradient during BP. The gradients need to be identified by which OP generats it and which needs it.}
\begin{center}
\begin{small}
\begin{adjustbox}{max width=\linewidth}
\begin{tabular}{lccccc}
\toprule
\multirow{2}{*}{Sub-DAG}  & OP & Required & Send & Required  & Send \\
&  nodes  &  Acti. (FP) & Acti. (FP)  & Grad. (BP) & Grad. (BP) \\
\bottomrule
\multirow{2}{*}{1} & Input,  & \multirow{2}{*}{Input} & \multirow{2}{*}{Conv}  & \multirow{2}{*}{Conv-Add} & \multirow{2}{*}{-} \\
&   Conv   &  &  &  &  \\
\midrule
\multirow{2}{*}{2} &  Tesor A & \multirow{2}{*}{-} & \multirow{2}{*}{ReLu}  & \multirow{2}{*}{ReLu-Add} & \multirow{2}{*}{-} \\
&  ReLu  &  &   &  & \\
\midrule
\multirow{2}{*}{3} & Label, & Conv,  & \multirow{2}{*}{-}  & \multirow{2}{*}{-}   & Conv-Add \\
&  Add, CE  & ReLu  &   &  & ReLu-Add \\
\bottomrule
\end{tabular}
\end{adjustbox}
\end{small}
\end{center}
\label{tab:subgraphs}
\end{table}

\subsection{DAG Partition}\label{sec:DAG_partition}

The whole training process can be divided into several procedures: FP, BP and Update. For a separate model, the FP and BP procedures cannot be completed by only one device. There must be communication of intermediate features (outputs from previous sub-models) and gradients between devices to enable FP and BP of subsequent sub-models. As shown in Figure~\ref{fig:DAG}, the whole computation and communication processes of FP and BP can be formalized as a DAG $\mathcal{G} = \langle \{ o^{i}\}_{i=1}^{n_o}, \{ o^{i},o^{j} \}  \rangle$, where $n_o$ is the total number of operators (OPs), node $o^{i}$ represents an operation $f^{i}$. The directed edge $(o^{i},o^{j})$ indicates that $o^{j}$ cannot start until $o^{i}$ has finished, and the outputs of $o^{i}$ will be sent to $o^{j}$. Table~\ref{tab:OPNodes} and Table~\ref{tab:subgraphs} show the attributes of these OP nodes and partitioned sub-DAGs. Next, we will illustrate how these definitions are used to help conduct FP, BP, and model Update.


 
\textbf{FP.} The FP data (edges in the DAG) can be identified by the OP names and the OP users. The outputs from the OP nodes will be sent to its OP users. And the args of an OP specify data from which OPs are required. When executing, as shown in Table~\ref{tab:subgraphs}, each sub-graph finds all the necessary inputs from other sub-graphs. Once all the inputs are available, the FP is launched and executed. The outputs of an OP are passed to its OP Users. When the executor detects that the OP Users are located on an external CompNode, it invokes communication functions to send the outputs to the corresponding external CompNodes. For example, the executor on CompNode 1 will transmit the outputs of the Conv OP to the Add OP on CompNode 3.


\textbf{BP.} In most cases, the BP edges are the reverse of the FP edges, except for the edges directed towards leaf nodes that do not require gradients, such as the Input and Label placeholders. The resources required for conducting BP are gradients from subsequent OPs. Similarly to the FP, the computed gradients are returned to their Arg Nodes. If the node is located on an external CompNode, the gradients are also communicated to that CompNode. For example, the executor will send the gradients on the outputs of the ReLu OP from CompNode 3 to CompNode 1 (these gradients are computed by the Add OP on CompNode 3).


\textbf{Update.} To support adaptive optimizers for different parametric OPs, users can define optimizers and corresponding hyperparameters in the configuration file. The broker assigns the appropriate optimizers to the target CompNode based on its assigned OPs. Once BPs are completed, i.e., the gradients are computed for the parametric OPs, the executor can utilize these optimizers to begin the optimization process.

\textbf{Message passing.} The executor possesses all the information about the sub-graph and can control the data flow between OP nodes. However, the executor does not have direct control over external OP nodes, which prevents it from moving data to those nodes. Therefore, for message passing between CompNodes, the send-side executor must determine which CompNode should receive the data, and the receive-side executor registers the message processing functions on the CompNode to store and process the required data. Table~\ref{tab:subgraphs} presents the attributes of subgraphs on CompNodes and their associated attributes, which are utilized for efficient and convenient message passing. The outputs of certain OPs can be quickly evaluated to determine whether they should be sent out or kept locally. When a CompNode receives data, it can be stored in the executor if it is part of the Outer Required Data, and vice versa.

\subsection{OP Data Structure}\label{sec:datastructure}
We design a uniform data structure that standardizes message exchanges between diverse operators and nodes, enhancing system coherence and reducing data processing complexity. Additionally, it facilitates automatic workload estimation during profiling by allowing for efficient analysis of standardized data, optimizing resource allocation, and DAG scheduling in real-time. Specifically, the OP data structure has the following attributes:

\textbf{Name} identifies the originating operation (OP) nodes that generate the data, providing traceability and aiding in debugging or optimization processes.

\textbf{OP users} lists the OP nodes that utilize this output as input, outlining data dependencies and aiding in the construction and optimization of the computational graph.

\textbf{Actual OP user} specifies the actual instances where the data is used, crucial for gradient calculations as it pinpoints the data's origin, thus ensuring accurate backpropagation and gradient application.

\textbf{Is\_loss} signifies whether the data in question pertains to output loss, assisting in loss tracking, and gradient updates across the network.

\textbf{Require\_grad} indicates if the data necessitates gradient calculations, central to efficient resource allocation since nodes can bypass unnecessary computations, focusing on gradient-relevant data.

\textbf{Local\_iter} and \textbf{micro\_batch} respectively reflect the current iteration and the number of micro-batches, essential for synchronizing operations on the same local iteration and micron batch in the realm of pipeline parallelism.

\textbf{Compress\_cfg} saves meta-information concerning data compression, encompassing details like the algorithm used, the compression ratio, and other pertinent hyper-parameters.

\subsection{Estimation of Computation and Communication}\label{sec:EstCompComm}
The decentralized computing system can be represented as a bidirectional graph $\mathcal{P}=\langle \{p^{i}\}_{i=1}^{n_p}, \{p^{i}, p^{j}\}\rangle$ with $n_p$ CompNodes. The GPU memory, computation speed and communication bandwidth of CompNodes $p$ are various.

The GPU memory is defined as $D^p$. The computation speed $S(p)$ of a CompNode is usually measured by floating-point operation per second (FLOPS).  However, the actual computation speed may not reach its peak performance ($S^*(p)$) provided by the original device information. This can be attributed to various factors, including the complexity of customized libraries like MKL~\citep{MKL} and cuDNN~\citep{chetlur2014cudnn}, as well as other parallel computations. To estimate the actual computation speed $S(p)$,  a regression-based scaling-down factor~\citep{qi2017paleo} $\lambda_p$ is used to estimate the actual computation speed $S(p)=\lambda_p S^*(p)$. Before scheduling, $\lambda_p$ is fitted by a short-time warmup profiling.

The communication cost between CompNodes $p^{i}$ and $p^{j}$ can be characterized by the alpha-beta model~\citep{MPICH,shi2019mg}: $T_{comm}^{ij}(M)=\alpha^{ij} + \beta^{ij}M$, where $\alpha^{ij}$ is the latency component, $\beta^{ij}$ the inverse of the communication speed, and $M$ the message size.

To model the overall consumed time ($T(f, p)$) of a specific operation $f$ on CompNode $p$, we incorporate a computation performance model~\citep{qi2017paleo} into our work:
\begin{equation}
    T(f, p) = \mathcal{R}(\text{Pa} (f)) + \mathcal{C}(f,p) + \mathcal{W}(f,p),
\end{equation}
in which $\mathcal{C}(f,p)$ represents the computation time, the $\mathcal{R}(\text{Pa} (f))$ represents the time of retriving data from parent nodes of $f$, $\mathcal{W}(f,p)$ the time of writing data into memory. Note that when $f$ and its parents $\text{Pa} (f)$ are located on different CompNodes, $\mathcal{R}(\text{Pa} (f))$ operation involves the communication and loading time between devices owning $f$ and $\text{Pa}(f)$. Otherwise, The communication won't happen between $f$ and $\text{Pa} (f)$.

\begin{table}[t!]
\centering
\caption{Notation Explanation}
\label{tab:notations}
\begin{tabular}{lr}
\toprule
\textbf{Notation} & \textbf{Explanation} \\
\midrule
$\mathcal{P}$& Group of CompNodes \\
$T_p$& One-time FP latency of device $p$ \\
$\mathcal{A}_p$& Subset of OPs assigned to CompNode $p$ \\
$\mathcal{C}_p$& Computation time on CompNode $p$ \\
$\mathcal{R}_p$& Communication time on CompNode $p$ \\
$P(f)$& The CompNode that has operator $f$ \\
$M_f$& Size of outputs from operator $f$ \\
$n_b$& Number of pipelined micro batches \\
$T(\mathcal{G}_{\mathcal{S}k})$& Time of one FP of the sub-graph $\mathcal{G}_{\mathcal{S}_k}$ \\
$T(\mathcal{G})_{lat}$& Time of one FP of the whole graph $\mathcal{G}$ \\
$T(\mathcal{G})_{n_b,pipe}$& Time of FP with pipelined batches \\
\bottomrule
\end{tabular}
\end{table}

We estimate the computation time as the floating-point oprations (FLOPs) counts of the operation $f$ divided by the actual computation speed: $\mathcal{C}(f, p)=\text{FLOPs}(f)/S(p)$.

Comparing with the computation and communication time, the IO time in $\mathcal{R}(\text{Pa} (f))$ and $\mathcal{W}(f,p)$ can be ignored. Thus, if $f$ and $\text{Pa} (f)$ are located on different devices, we estimate the data retriving time as $\mathcal{R}(\text{Pa} (f))=T_{comm}^{ij}(M)$, in which $i$ and $j$ are CompNodes that $f$ and $\text{Pa} (f)$ locate at.

\subsection{Estimation of Workloads and Througput}\label{sec:throughput}

A DAG of the neural network has operators that can be executed sequentially or in parallel. As a result, the executing time $T(\mathcal{G}_{\mathcal{S}_k})$ of a sub-graph $\mathcal{G}_{\mathcal{S}_k}$ on CompNode $p$ is within the range $[\text{max}_i T(f^{i},p), \sum^{i} T(f^{i},p) ]_{i\in\mathcal{S}_k}$.

We consider training a LLM with a group of CompNodes, where each CompNode is responsible for processing a sub-DAG of the LLM. Given the hardware performance obtained in \S~\ref{sec:EstCompComm} and partitioned sub-DAGs $\{\mathcal{S}_k\}_{k\in\mathcal{K}}$ ($\mathcal{K}$ is the set of sub-DAGs), and a group of CompNodes $\mathcal{P}$, 

The computation time of each device is approximated by $T_p=\sum_{k\in\mathcal{A}_p} T(\mathcal{G}_{\mathcal{S}_k})$. Because the tensor parallelism would introduce much more extra communication costs between devices,  we simply consider the pipeline parallelism in this paper for now. Thus, for most famous deep models, sub-DAGs are sequentially executed. Because the read and write time in the local device is little comparing with the training and communication, we can remove $\mathcal{R}(\text{Pa}(f))$ if $\text{Pa}(f)$ and $\mathcal{W}(f,p)$ local on the same $p$ with $f$.

Then we can formalize the latency of FP process as:
\begin{small}
\begin{align}
   T(\mathcal{G})_{lat}  & = \sum_p^{p\in\mathcal{P}} T_p = \sum_p^{p\in\mathcal{P}} \sum_{k\in\mathcal{A}_p} T(\mathcal{G}_{\mathcal{S}_k})
   = \sum_p^{p\in\mathcal{P}} (\mathcal{C}_p + \mathcal{R}_p) \notag \\ 
    \mathcal{C}_p  & = \sum_{k\in\mathcal{A}_p} \sum_{f \in\mathcal{S}_k} \mathcal{C}(f,p) \notag \\
    \mathcal{R}_p  & = \sum_{k\in\mathcal{A}_p} \sum_{ \substack{ f \in\mathcal{S}_k, \\ P(f) \neq P(\text{Pa}(f))}}  \mathcal{R}(\text{Pa}(f)),
    \label{eq:time_latency}
\end{align}
\end{small}
in which $P(f)$ maps operator $f$ to its located CompNode $p$, $M_f$ represents the size of outputs from operator $f$, $\mathcal{R}(\text{Pa}(f)))= T_{comm}^{P(f,\text(Pa)(f)) }(M_f)$. When we apply pipelining different batches during training~\citep{huang2019gpipe}, some computation time and communication time can be overlapped. Considering $n_b$ batches are pipelined, the time cost of processing them is:
\begin{small}
\begin{equation}\label{eq:time_pipe}
   T(\mathcal{G})_{n_b,pipe}  = \sum_p^{p\in\mathcal{P}} (\mathcal{C}_p + \mathcal{R}_p) + (n_b - 1) \max_{p\in\mathcal{P}} (\mathcal{C}_p, \mathcal{R}_p).
\end{equation}    
\end{small}
Thus, the throughput of processing samples should be:
\begin{equation}\label{eq:throughput}
    \phi = N_s / T(\mathcal{G})_{n_b,pipe},
\end{equation}
in which $n_s$ is the batch size of the mini-batch sampling.

\section{OP-Fence}\label{sec:opfence}
In our decentralized system, efficient scheduling of operator nodes within the forward propagation DAGs is paramount. Such scheduling directly impacts computational load balance, ensuring that resources are allocated judiciously to avoid overutilization of certain nodes while underutilizing others. Equally critical is the memory balance; correct allocation prevents memory saturation and mitigates the risks of bottlenecks or performance degradation as shown in \S~\ref{sec:throughput}. Furthermore, in a distributed computing environment, communication balance can't be understated. Proper scheduling optimizes data exchanges, ensuring seamless synchronization across nodes. In essence, meticulous scheduling in our system is analogous to the fine-tuning of a high-performance computing cluster, where every decision has cascading implications on overall system efficiency and throughput.

\begin{figure}[t!]
    \subfigtopskip=2pt
    \setlength{\belowdisplayskip}{2pt}
    \setlength{\abovedisplayskip}{-5pt}
    \subfigbottomskip=2pt
    \subfigcapskip=1pt
   \centering
    {\includegraphics[width=0.89\linewidth]{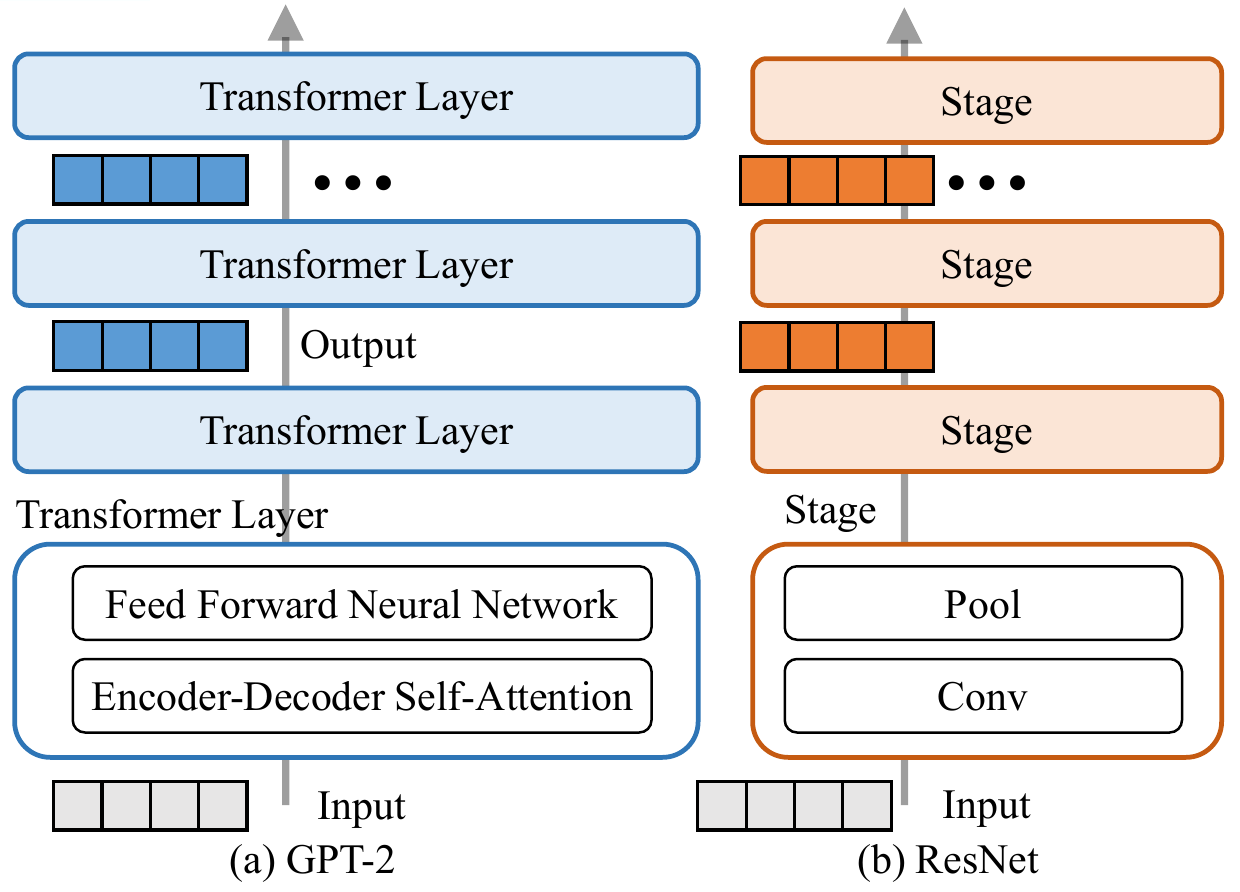}}
    \caption{GPT-2 and ResNet Architectures.}
    \label{fig:DNNs}
\end{figure}

\begin{figure}[t!]
    \subfigtopskip=2pt
    \setlength{\belowdisplayskip}{2pt}
    \setlength{\abovedisplayskip}{-5pt}
    \subfigbottomskip=2pt
    \subfigcapskip=1pt
   \centering
    {\includegraphics[width=0.99\linewidth]{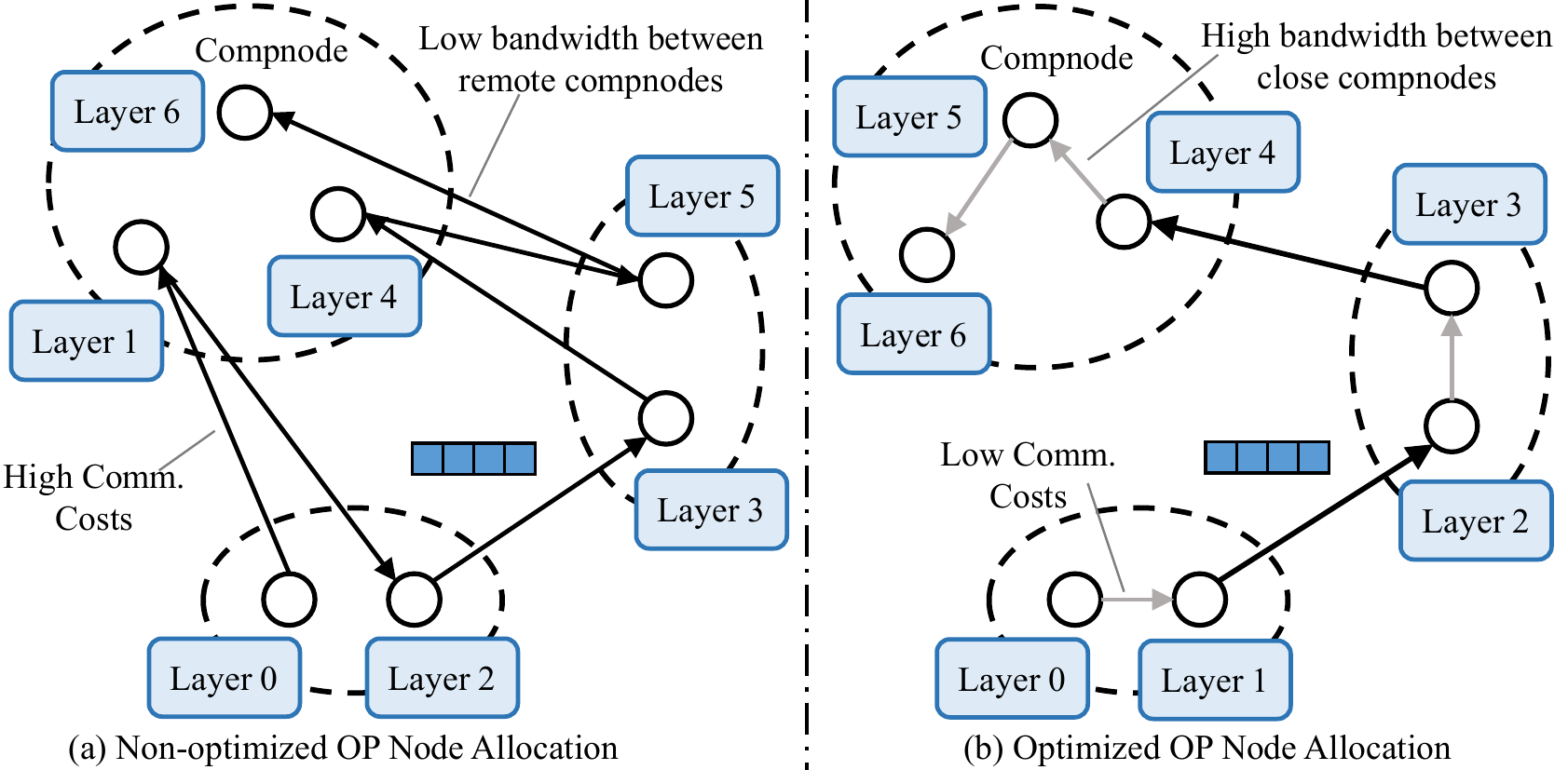}}
    \caption{Adjacent CompNodes have higher communication bandwidth. Arrows show the FP direction in the DAG. Different allocations of different OP Nodes lead to significant difference of communication costs.}
    \label{fig:NodeCluster}
\end{figure}

Because the FP and BP DAGs have the same structure, only with reverted directions, we only conduct partition and scheduling on FP DAG in the rest of the paper for simplicity. The throughput optimization problem in our decentralized training system can be formulated as following:
\begin{align}    
    \min_{\mathcal{A}} \ \ &  \Tilde{T}(\mathcal{G})_{n_b,pipe}  \\ \label{eq:schedule}
s.t. \ \  &  D_{gpu}^p \geq  \sum_{k\in\mathcal{A}_p}  D_{gpu}(\mathcal{G}_{\mathcal{S}_k}),  \\ \notag 
 \ \  &  D_{cpu}^p \geq  \sum_{k\in\mathcal{A}_p} D_{cpu}(\mathcal{G}_{\mathcal{S}_k}), \\ \notag 
 \ \  &  D_{disk}^p \geq  \sum_{k\in\mathcal{A}_p}  D_{disk}(\mathcal{G}_{\mathcal{S}_k}).  \notag
\end{align}

Because this optimization problem is too complicated to be solved in a closed form, we assume that each device has enough CPU and disk memory to load and run sub-DAGs, which is a reasonable assumption in real-world. Then, we use the following observations to design our scheduling algorithm.

\textbf{Observation 1: degree of the DAGs of deep learning model is usually small (less than 2).} As shown in Figure~\ref{fig:DNNs}, the architecture of DNNs is often characterized by layers that are sequentially stacked one by one, akin to a chain~\citep{Resnet,gpt2,gpt3}. A layer (or operator node) usually receives input from only the preceding layer and sends its output to the subsequent layer, rather than branching out or converging from multiple layers, as you would see in a tree structure or a complex city roadmap. Thus, the degree of these DAGs is typically small.

\textbf{Observation 2: network locality.} 
In decentralized networks, bandwidth among participants varies, often influenced by geography. Research in computer networking indicates that those in close proximity usually enjoy higher bandwidth, while distant interactions face bandwidth constraints~\citep{Gulati1993,locality}. This pattern suggests the presence of high-bandwidth clusters within the broader network graph, highlighting areas of efficient data transfer. Recognizing these clusters is key to optimizing data exchanges in collaborative endeavors like decentralized training.

Leveraging these observations, in \ourscheduler{}, we detect the high-bandwidth clusters in computing devices using Louvain algorithm~\citep{blondel2008fast}. Each device communicates with other devices in the same cluster with a high bandwidth. Then, \ourscheduler{} makes that each cluster is allocated operators that form a connected sub-graph, thereby minimizing the necessity for data to traverse low-bandwidth paths. Figure~\ref{fig:NodeCluster} illustrates a comparison between an unoptimized scenario and an optimized one, highlighting the efficiency achieved through this deliberate structuring.

\section{AdaTopK Compressor}\label{sec:adaTopk-AG}

\subsection{Basic Communication Compression}\label{sec:basiccompression}
\begin{figure}[t!]
\unishrink{}
    \subfigtopskip=2pt
    \setlength{\belowdisplayskip}{2pt}
    \setlength{\abovedisplayskip}{-5pt}
    \subfigbottomskip=2pt
    \subfigcapskip=1pt
   \centering
    {\includegraphics[width=0.99\linewidth]{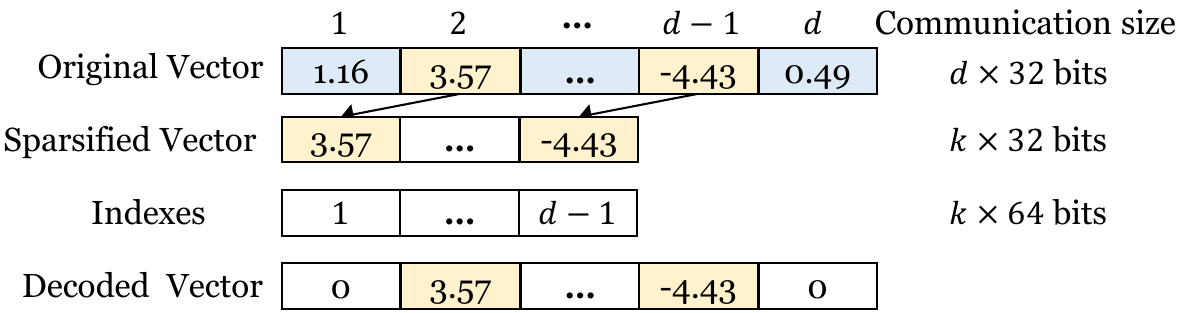}}
\unishrink{}
    \caption{Top-K sparsification of a vector with dimension $d$. The $k$ largest absolute values are chosen to communicated, other elements are left as 0 when decompressing. Thus, the communication size is reduced from $32d$ to $64k$.}
    \label{fig:compression}
\unishrink{}
\end{figure}

The overlap between communication and computation in training DNNs underscores the need for communication efficiency to fully harness the computational potential of distributed systems. Any delay or inefficiency in communication can hinder the simultaneous processing advantages, turning potential speed-ups into idle waiting times. Therefore, ensuring communication efficiency is pivotal for accelerating training time and making the most of distributed resources.
Quantization~\citep{ProbroundingNN,DorefaNet,QSGD,TernGrad,zhang2023evaluation} uses lower bits to represent data originally represented by 32 bits on each dimension of the transmitted gradient. Sparsification~\citep{Hispeed,EffiUseofLimitMemory,li2022on,zhang2023evaluation,pmlr-v119-fu20c,ACGC,dong2024pruner,tang2020communication,tang2022gossipfl} means to pick up a part of values in the gradients or model parameters to communication. Figure~\ref{fig:compression} shows the Top-K sparsification of a vector, which is a widely used method in data parallel distributed training~\citep{9155269,tang2020communication}.

\begin{figure}[t!]
\begin{minted}[breaklines, frame=lines,fontsize=\scriptsize]{python}
import torch
import fusionai
from fusionai.nn.op_node_cfg import Dataset_OP_Node_cfg
from fusionai.nn.op_node_cfg import Module_OP_Node_cfg
from fusionai.nn.op_graph_cfg import OP_Graph_cfg

@fusionai.TORCH_MODULES.register("CustomReLU")
class CustomReLU(nn.Module):
    def __init__(self):
        super(CustomReLU, self).__init__()
    def forward(self, input):
        return torch.where(input>-1, input, torch.zeros_like(input))

class TwoLayerNet(nn.Module):
    def __init__(self, input_size, hidden_size, output_size):
        super(TwoLayerNet, self).__init__()
        # build input named as x.
        x = Dataset_OP_Node_cfg(name="x", type="input")
        self.add_op_node(x, add_module=False)  # add x into DAG.
        conv1 = Module_OP_Node_cfg(name="conv1", type="Conv2d",
                    kwargs=dict(kernel_size=3, padding=1,
                    stride=1, in_channels=3, out_channels=64))
        conv1.add_args(x)        # add x as an input into conv1.
        self.add_op_node(conv1)  # add conv1 into DAG.
        myrelu = Module_OP_Node_cfg(name="myrelu", type="CustomReLU")
        myrelu.add_args(conv1)   # add conv1 as an input into myrelu.
        self.add_op_node(myrelu) # add myrelu into DAG.
\end{minted}
\captionof{figure}{An example of a CNN defined in \oursys{}.}
\label{fig:CodeSnippet}
\end{figure}

\subsection{Adaptive Communication Compression}\label{sec:adacompression}
Although communication compression can uniformly reduce the communication time between each node, the model convergence might be affected. To strike a better trade-off between training convergence and communication time, a better way is to compress data between those nodes with lowest bandwidth. To this end, given a compression ratio by users, we calculate the new compression ratio $r$ between nodes according to the following equation:
\begin{equation}\label{eq:adacompress}
    r_{i} = \text{max}(1, 3r \times \mathcal{R}_i /  \mathop{max}_{p\in\mathcal{P}}(\mathcal{R}_p)),
\end{equation}
in which the new compression ratio is decided by the estimated original communication time, the coefficient 3 comes from that the Top-K compression requires to send both compressed values (float32) and the indexes (int64). From another view of such an adaptive compression, the time cost of the pipeline becomes:
\begin{equation}\label{eq:time_pipe_compress}
\small
   \Tilde{T}(\mathcal{G})_{n_b,pipe}  \simeq \sum_p^{p\in\mathcal{P}} (\mathcal{C}_p + 3\mathcal{R}_p / r_{i}) + 3(n_b - 1) \max_{p\in\mathcal{P}} (\mathcal{C}_p, \mathcal{R}_p)/r,
\end{equation}
which implies that when we increase the $n_b$ to a large value, the decrease speed of the bottleneck term with the $r$ is not affected by the adaptive compression algorithm. This would be very useful in the high $n_b$ pipeline parallel training.

\section{Implementation}\label{sec:implementation}

\oursys{} is independent of ML frameworks. In the current version, \oursys{} is built upon Python and libraries including PyTorch~\citep{torch} as the ML engine, torchtext~\citep{torch} and HuggingFace Transformers~\citep{wolf2019huggingface} as NLP toolkits. In future versions, \textbf{} will support other ML frameworks like TensorFlow~\citep{abadi2016tensorflow}, MXNet~\citep{chen2015mxnet}, PaddlePaddle~\citep{paddle} and so on. The core codes of \oursys{} (excluding models, datasets and configurations) are totally 11.3K lines in Python and around 0.5K lines in C++.

\textbf{Communication.} In current version of our system, we utilize the N2N tool and MPI collective communication to build the P2P connection between GPUs. We will implement the P2P connection without N2N in the future version. And we will implement a more efficient communication manager to let GPUs choose fast links, like NCCL connections in a same cluster.

\textbf{Compression.} We implement a TopK sparsification library at Cuda level that is faster than PyTorch TopK operation, to reduce the compression time.

\textbf{Model definition interface.} \oursys{} supports customized model definition. Figure~\ref{fig:CodeSnippet} shows how to define a two-layer convolutional neural network (CNN) through a few lines of codes. In this example, the ML framework is chosen as PyTorch, and users use a PyTorch Conv layer and a customized ReLu activation function to build a simple CNN.

\section{Experimental Studies}\label{sec:exp}

\begin{figure*}[h]
    \subfigbottomskip=-1pt
    \subfigcapskip=1pt
    \centering
     \subfigure[ResNet18 with CIFAR-10.]{\includegraphics[width=0.32\textwidth]{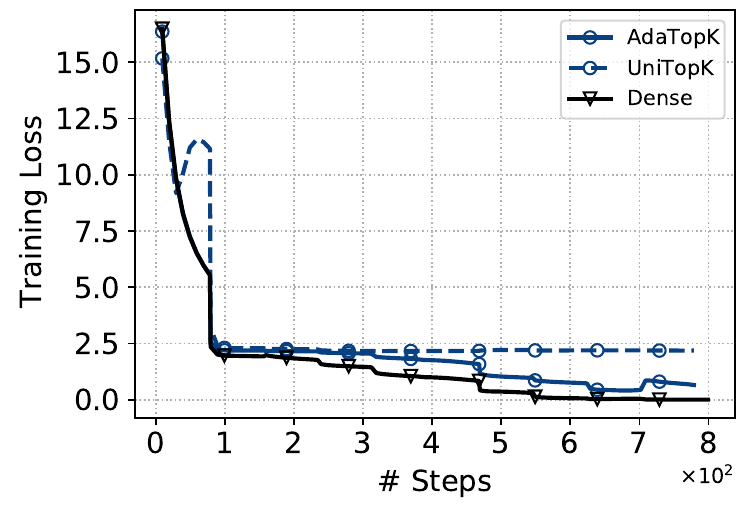}}
     \subfigure[ResNet101 with Tiny-ImageNet.]{\includegraphics[width=0.32\textwidth]{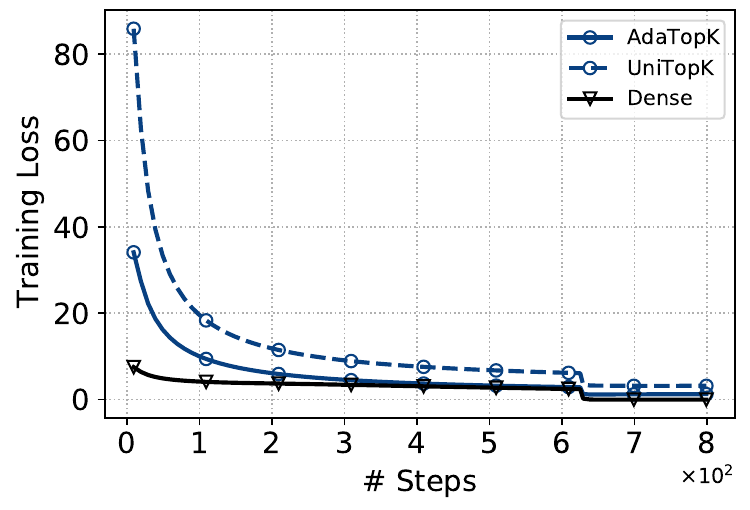}}
     \subfigure[GPT2-XL with Wikitext-2.]{\includegraphics[width=0.32\textwidth]{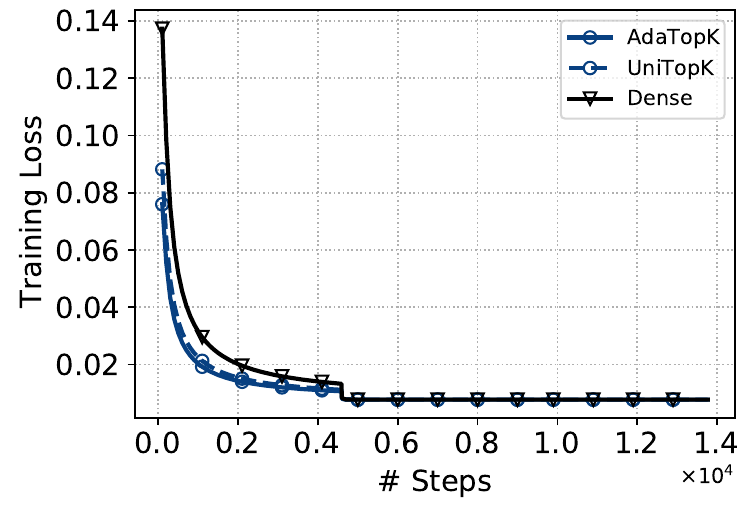}}
    \caption{Training loss with different model architectures and datasets. ``Dense'' means training without compression. All compression ratio is set as 100.}
    \label{fig:Convergence}
\unishrink{}
\end{figure*}

\subsection{Experimental Setup}\label{sec:exp_setup}

\begin{figure}[h]
    \subfigbottomskip=-1pt
    \subfigcapskip=1pt
    \centering
     \subfigure[Latency (seconds).]{\includegraphics[width=0.49\linewidth]{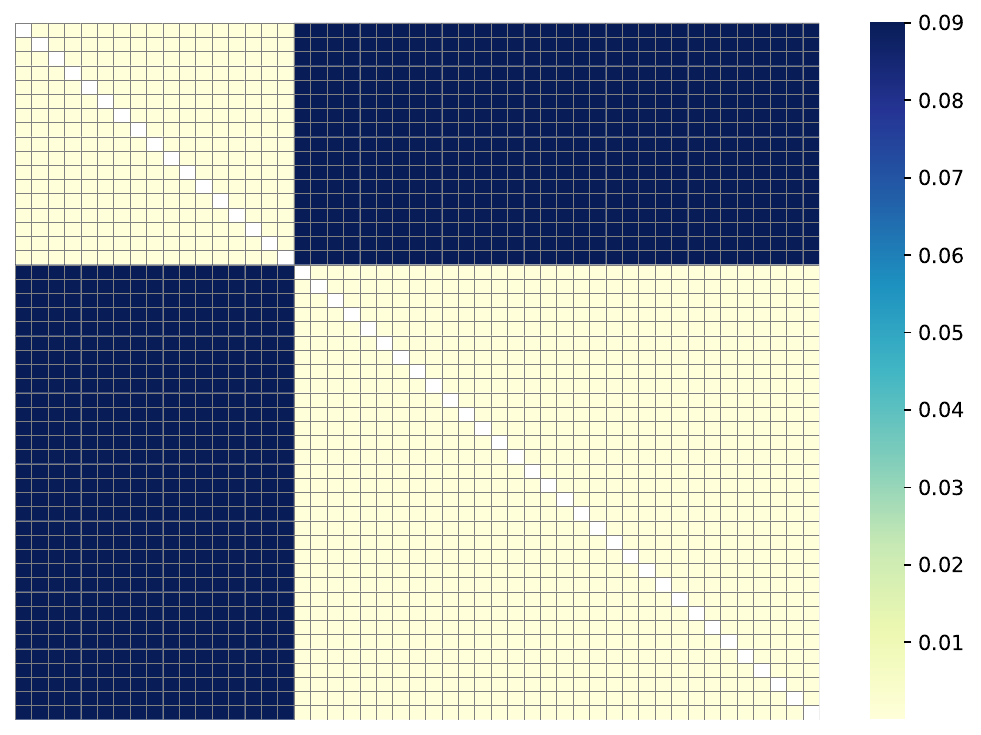}}
     \subfigure[Bandwidth (MB/s).]{\includegraphics[width=0.49\linewidth]{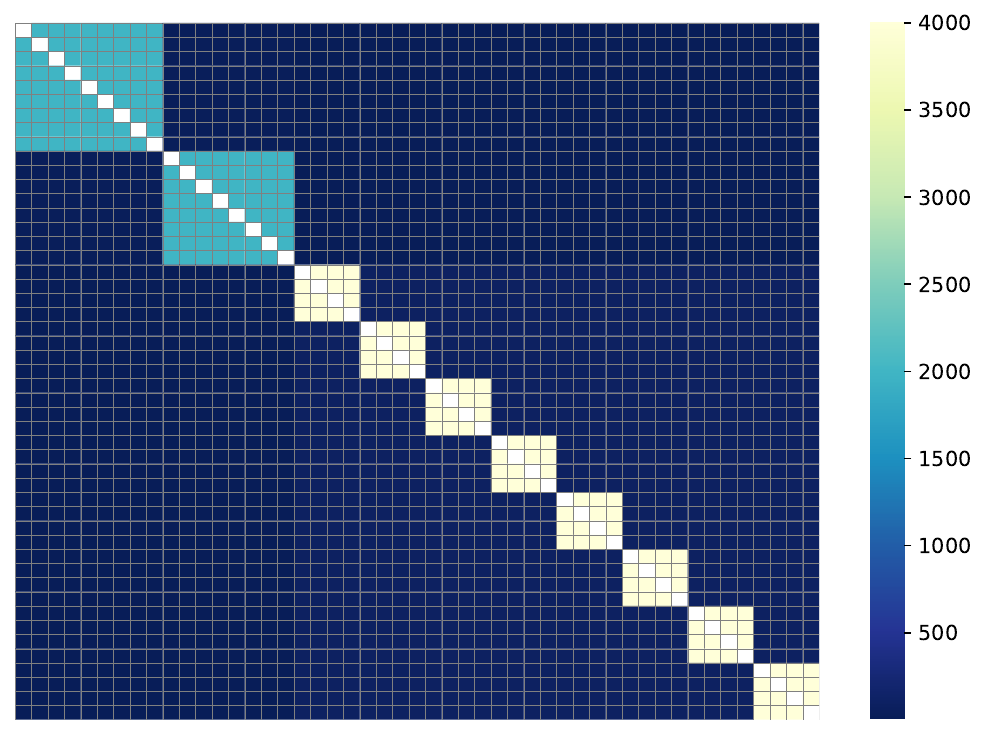}}
    \caption{Latency and bandwidth of 48 CompNodes in all testbeds.}
    \label{fig:communication_params}
\unishrink{}
\end{figure}

\textbf{Testbeds.} We use two different clusters in our testbeds. Cluster A has 2 GPU machines (called node 1-2), each with 8 RTX 4090s and equipped with Ubuntu 20.04.5. PyTorch 2.1.0, Cuda 12.1.
Cluster B has 8 GPU machines (called node 3-10), each with 4 RTX 2080s, and runs with Ubuntu 20.04.6 and the software environment includes CUDA 12.1, PyTorch 2.2.0. The communication latency and bandwidth between GPUs are heterogeneous as shown in Figure~\ref{fig:communication_params}.

We design four testbeds as shown in Table~\ref{tab:testbeds}. Each GPU is regarded as a compute provider who joins the decentralized training. Note that the communication between GPUs in one same machine does not use NCCL for simulating lower communication bandwidth in realistic world.

\begin{table}[h]
    \centering
    \caption{Number of GPUs and GPU locations in different testbeds. $n\times k$ means utlizing $n$ nodes and each node with $k$ GPUs in according cluster.}
    \begin{tabular}{cccr}
    \toprule
    \textbf{Testbed} \textbackslash~ Cluster & A & B & total \\
    \midrule
    \textbf{1} & $1\times 8$ & $4\times 4$ & 24 \\
    \textbf{2} & $2\times 8$ & $8\times 4$ & 48 \\
    \bottomrule
    \end{tabular}
    \label{tab:testbeds}
\end{table}

\subsection{End-to-End System Comparisons}






\begin{table}[h]
    \centering
    \caption{Benchmark settings and information of different models.}
    \resizebox{\linewidth}{!}{
    \begin{tabular}{lcccc}
    \toprule
    Model & \# Parameters & Input size & Batch size  & \# micro batches  \\
    \midrule
    ResNet18 & 11.69B & $3\times 32\times 32$ & 128 & 5 \\
    ResNet101 & 44.55B & $3\times 64\times 64$ & 32 & 5 \\
    GPT2-XL & 175B & $30200\times 1024 $ & 3 & 2 \\
    \bottomrule
    \end{tabular}
    }
    \label{tab:modelsettings}
\end{table}

\textbf{Workloads.} We use the large real-world DNN models including well-known computer vision model (ResNet18 and ResNet101~\citep{Resnet}), and the NLP model (GPT2-XL~\citep{gpt2}) following the literature~\citep{NEURIPS2022_7a43b8eb,yuandecentralized,258953,EfficientSparseCollecComm}. The benchmark settings of three different models are shown in Table~\ref{tab:modelsettings}. And we validate the convergence of training ResNet18 on CIFAR-10, ResNet101 on Tiny-ImageNet~\citep{le2015tiny}, GPT2-XL on WikiText-2~\citep{gpt2} with our compression algorithm. As LLM like GPT2-XL is our main focus, we measure its latency four different testbeds.

\textbf{Baselines.} We use collaborative training without compression as the basic baseline. For compression algorithm, we use naive Top-k compression as baseline. For scheduling algorithm, we use two baselines: 1) equal number of modules, which equally assigns the same number of user-defined modules on computing nodes; 2) equal computation costs, which firstly partitions user-defined modules with load balance and assign each computing node with on partition. 

\textbf{Performance metrics.} We use the averaged time costs as the latency of training one iteration (including processing all micro batches). To validate the convergence of training, we also show the curve of training loss of ResNet-18, ResNet-101 and GPT2-XL in experimental results.



\subsection{Convergence of Training}
Figure~\ref{fig:Convergence} shows the convergence curve of training three different DNNs with \oursys{}. For both ResNet-18 and ResNet-101, the uniform TopK compression affects convergence severely. while our presented adaptive TopK exists little convergence gap to no-compression training. This demonstrates that although uniform TopK can significantly reduce communication time by compressing communicated data between each nodes, this negatively influences the training, due to too much lost information.

Interestingly, training GPT2-XL on Wikitext-2 with both uniform and adaptive TopK compression does not influence training convergence, but improve the convergence reversely. Here are several possible reasons why this happens and also why adaptive TopK influences training ResNet little: 

(1) Noise Reduction: compression method like quantization or sparsification, can effectively act as a form of noise reduction~\citep{LeCun2012}. By only keeping the most significant parts of the gradients (e.g., the largest values in TopK sparsification), the stochastic noise that is inherent in the gradient estimates (due to mini-batch gradient descent) can be reduced~\citep{neelakantan2015adding}. This noise reduction can make the optimization process smoother and more stable, potentially leading to better convergence. 

(2) Implicit Regularization: Compression can also serve as a form of implicit regularization~\citep{wilson2017marginal,zhang2021understanding}. By compressing the gradients, you're essentially limiting the capacity of the updates the model can make at each step, which can prevent overfitting. This is somewhat analogous to techniques like dropout~\citep{wager2013dropout} or weight decay~\citep{zhang2018three}, which are known to improve generalization and thus help convergence in terms of better validation loss.

\subsection{System performance}

\begin{figure}[h]
    \subfigbottomskip=-1pt
    \subfigcapskip=1pt
    \centering
     \subfigure[Testbed1.]{\includegraphics[width=0.48\linewidth]{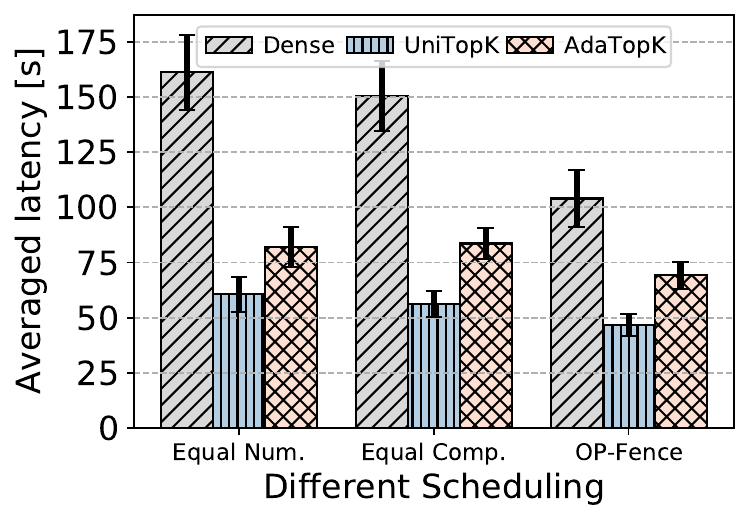}}
     \subfigure[Testbed2.]{\includegraphics[width=0.48\linewidth]{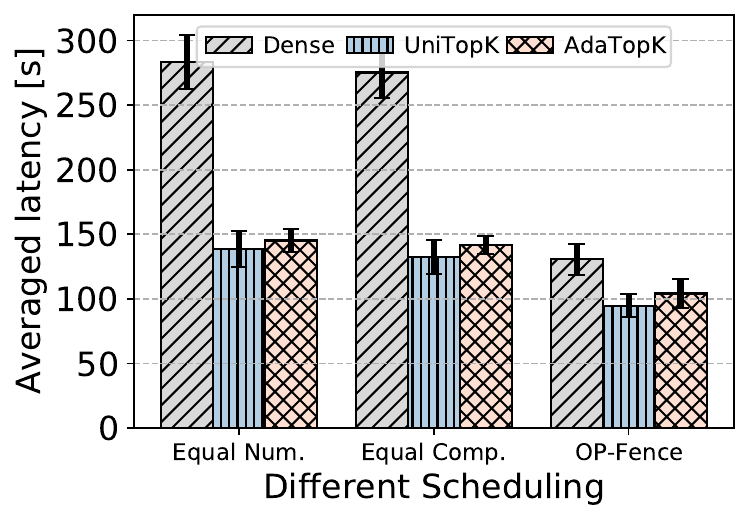}}
    \caption{Averaged latency of training one iteration on different testbeds with different scheduling and compression algorithms. All compression algorithms use 100 compression ratios. The actual compressed data is $33.3\times$ less than original data (values with float32 and indexes with int64).}
    \label{fig:latency}
\unishrink{}
\end{figure}

Figure~\ref{fig:latency} shows the latency of training with different scheduling and compress methods. In terms of scheduling, the naive partition method - equal-number partition - shows the highest latency. According to our profiling, the FP and BP time in the GPT2-XL consumes less than 0.5 seconds, while the intermediate features occupy around 20 MB, leading to 20 seconds to communicate with the 1MB/s bandwidth. Thus, the bottleneck is the communication instead of the computation, the partitioning based on the equal computation costs reduces the latency a little. 

In terms of compression, training without compression consumes the most time across all methods. Both uniform and adaptive TopK can reduce the latency a lot. Because the uniform TopK reduce communication size between each pair of nodes, it enjoys lower latency than adaptive TopK. However, as shown in Figure~\ref{fig:Convergence}, uniform TopK may result in significantly harmed convergence, thus enlarging the training time inversely. Also, because the bandwidth between nodes is heterogenous, uniform TopK cannot obtain lower latency than adaptive TopK with a large gap.



\subsection{Effect of Compression Ratio}
\begin{figure}[t!]
\unishrink{}
    \subfigtopskip=2pt
    \setlength{\belowdisplayskip}{2pt}
    \setlength{\abovedisplayskip}{-5pt}
    \subfigbottomskip=2pt
    \subfigcapskip=1pt
   \centering
    {\includegraphics[width=0.7\linewidth]{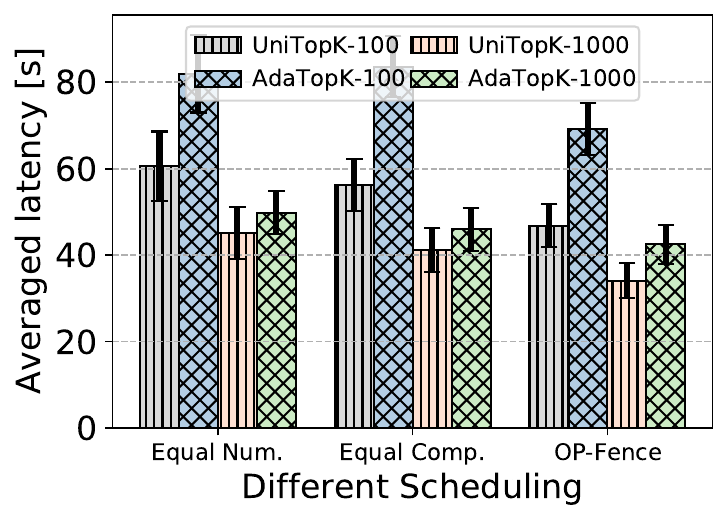}}
    \caption{Training with different compression ratios of 100 and 1000.}
    \label{fig:DifCompressRatio}
\unishrink{}
\end{figure}

Figure~\ref{fig:DifCompressRatio} shows that the increased compression ratio of 1000 may not reduce the latency for 10$\times$ less than ratio of 1000. The reason maybe that the training latency comes from other terms like latency per message ($\alpha$) and inner scheduling become the bottleneck.

\section{Limitations}\label{sec:limitations}
\textbf{P2P communication performance.} The communication performance of our current version of \oursys{} is limited by the N2N tool and MPI collective communication. Specifically, the collective communication is build upon the virtual IP addresses constructed by the N2N server, which exists in another remote area that need to be communicated over Internet. Thus, all P2P communication need to be processed by this remote N2N server, significantly limiting the utility of bandwidth resources of P2P communication.

\textbf{Network stability.} Our current system assumes a certain level of network stability, which might not be the case in all real-world scenarios, especially over the Internet. During each training iteration, the loss of communication packet, various latency, connection interruptions, hardware fault and other problems might severely affect the training.

\textbf{Economic and practical feasibility.} This paper proposes a method for utilizing geo-distributed GPUs, but it doesn't thoroughly investigate the economic aspects of such a setup. The costs associated with data transmission, especially in geographically diverse setups, and the incentive mechanisms of getting multiple entities to contribute resources, are not deeply analyzed.

Despite the presence of these unaddressed challenges within our current framework, this paper crucially establishes a proof-of-concept for the geo-distributed training of DNNs and LLMs. Our research underscores the viability of this training system, breaking ground for a new paradigm in distributed ML. Furthermore, the results we've obtained signal considerable potential for enhancements and optimizations in future iterations. 




\section{Related Works}\label{sec:related}

\subsection{Decentralized DL}\label{sec:opentraining}
Peer-to-peer computing~\citep{DONet,diskin2021distributed} and scientific computing~\citep{5719609} have previously explored the potential of utilizing massive personal and edge devices. Walle~\citep{Walle} is a device-cloud ML system that distributes ML tasks with small models between devices, while MOSAIC~\citep{mosaic} and JALAD~\citep{8645013} focus on efficient inference and only verify their systems on small models and datasets. ComAI~\citep{9796769} proposes a cloud-edge collaborative inference framework.

Many current edge works focus on efficient inference, federated or transfer learning with multiple small models~\citep{EIPaving,9628184,tang2022gossipfl}, whilst our system concentrates on exploiting massive devices for training DNNs. DeDLOC~\citep{DeDLOC} implements volunteer computing to train ALBERT~\citep{Lan2020ALBERT} with DP because ALBERT is small enough to be loaded onto a single Nvidia V100.

SWARM~\citep{ryabinin2023swarm}, Learning@home~\citep{DecentMOE}, CocktailSGD~\citep{cocktailSGD}, AQ-SGD~\citep{NEURIPS2022_7a43b8eb} and DT-FM~\citep{yuandecentralized} are the first to use low-bandwidth connected devices to train larger models such as GPT-2 and large Transformers. However, they do not consider the heterogeneity of consumer-level GPUs, the variability of collaboration, the compatibility of software and hardware, or the generality of DL tasks.


\subsection{Edge DL}

Edge devices have limited resources in terms of memory, computational power, communication, and energy, which encourages memory-efficient and data-efficient training and inference. Low-precision training~\citep{NEURIPS2018_335d3d1c,cambier2020shifted} and trading memory for computation~\citep{gruslys2016memoryefficient,chen2016training} are designed for high-throughput cloud training of large models with high storage requirements, but may not be suitable for low-storage edge devices. This work aims to provide scalability for large-scale dataset and model training and inference on edge devices by partitioning the workload of large model training into multiple subtasks, adapting to different device topologies.

\subsection{Incentive Mechanism}\label{sec:incentive}

Incentive mechanisms serve as economic catalysts to facilitate decentralized training. Some previous schemes like auction, contract, Stackelberg game, etc. can be employed to incentivize some selfish and rational clients to participate in this cooperative learning~\citep{10.1145/2764468.2764525,10.1145/3589303,kwon2021beta}. Compared with previous incentive schemes in other similar scenarios, there are some special challenges and considerations in the mechanisms design. (1) The property of online training indicates that the arrival and departure time is unknown and varies drastically for different clients. This online property is in line with asynchronous training but renders previous one-round incentive schemes ineffective. (2) Some other alternative choices should be considered for clients like bitcoin mining, client-assisted contributions, etc.,  which can also bring some economical benefits to their participation. In other words, incentive design should attract clients to join in decentralized learning from many candidate choices. (3) The incentive scheme should be robust and resilient to some malicious clients which contribute nothing but endeavor to get large paybacks.

\subsection{Graph Partition}\label{sec:graphpartition}
The load balancing (to reduce computation or storage costs) and minimizing cuts (to reduce communication costs) of graph partitioning is called balanced graph partitioning problem. Optimizing these two objectives simultaneously is known as an NP-hard problem~\citep{GraphEdgePartition,doi:10.1137/S1064827595287997}. METIS~\citep{doi:10.1137/S1064827595287997} starts with a random initial partition of the graph and optimizes the partition by continuously swapping vertices and edges between subgraphs. The partitioning algorithm determines the placement of the incoming vertices or edges based on a set of heuristic rules. ~\citep{7161590} adds bandwidth-aware features to the traditional partitioning algorithm, enabling it to compute partitioning schemes based on the imbalanced communication capabilities between nodes. ~\citep{6975163} considers a distributed computing environment made up of multi-core processors and can design partitioning schemes based on the computational capacity of each node and the communication bandwidth between each pair of nodes. ~\citep{10023963} partitions a CNN onto different edge devices to complete training.

\section{Conclusion}\label{sec:conclusion}

In this paper,  we present \oursys{}, a decentralized system for training 
DNNs with geo-distributed GPUs, designed to mitigate computational demands and enhance data privacy. Our system's core, \ourDAG{}, allows for general model customization, remote automatic differentiation, and supporting diverse software environments prevalent in decentralized computing scenarios. Through strategic partitioning of LLMs, \oursys{} enables collaborative training across varied computational nodes, democratizing LLM development beyond high-resource entities. For heightened efficiency, \oursys{} incorporates \ourscheduler{} for optimal resource allocation and \ourcompressor{} for critical data compression, ensuring robust performance without compromising training integrity.

In summary, \oursys{} revolutionizes DNN training by decentralizing the process, thereby breaking barriers of geography, technology, and cost. This approach paves the way for more inclusive, scalable, and efficient machine learning endeavors, potentially transforming LLMs from elite commodities to accessible global assets.

Building upon the innovative foundation of \oursys{}, there is substantial potential for future work to expand its capabilities and performance in aspects including advanced compression algorithms, robustness and security enhancements, dynamic resource allocation and more.


\bibliography{cites}

\appendix

\end{document}